# Graphene 'Battery' made of Low Cost Reduced Graphene Oxide

By Zihan Xu [1*]

1   Carbon Source Technology, LLC
    Shenzhen, Guangdong, China
*   zihan.xuu@gmail.com, RDzxu@csourcetech.com

Graphene can collect energy from the ambient heat and convert it to electricity[1], which makes it an ideal candidate for the fabrication of self-powered devices[2,3]. However, it was suffering several disadvantages, such as high cost for the fabrication of high quality graphene. The cost of gold electrod was also very high when considering making it a large scale one. Besides, the substrate of the graphene device, means $SiO_2$/ Silicon wafer, also limits the application of the device.

Here, we demonstrated that graphene "battery" can be made of low cost Reduced Graphene Oxides (RGO). We also replaced the gold electrode with graphite, which cuts the cost a lot. Besides, glass, instead of silicon dioxide/ silicon substrate was used in this work, showing the potential of making this product much cheaper and makes the processes much simpler.

Graphene Oxide (GO) was prepared with the modified Hummer's Method [3,4]. 5 g flake graphite (size: 1mm, purity:99.8%) and 5 g $NaNO_3$ was put into a beaker, then 150mL 97% sulfuric acid was added. The mixture was stirred for 2 hours at 50°C. Then it was set into ice-bath until its temperature dropped below 10°C. 15 g KMnO4 was added slowly into the beaker and the temperature of the mixture was kept below 10 °C. The reaction lasted for 2 hours while being stirred. Then 40mL deionized (DI) water was added to the beaker slowly, the speed was kept at about 4mL/ min and the beaker was kept in ice bath. After the 40mL water was added, the water bath was heated to 35°C and was kept at this temperature for 1 hour, and then it was heated to 85°C for 30mins. 300mL DI water was added into the mixture and the temperature dropped to 50oC. 5mL $H_2O_2$ (30%) was dropped to the solution to abort the reaction.

The solution was first centrifuged at 2000rpm for 5mins to remove the unreacted graphite and then at 8000rpm for 15mins to remove the small flakes. Then the remained colloid GO was washed by a centrifuge at 10000 rpm for 5 minutes with 1% HCl for three times and then with DI water for 3 times. The remained GO colloid was dispersed in ethanol to about 3% by weight. The dispersed GO colloidal solution was treated with ultrasonic cleaner (30°C, 50W) for 15mins to reduce the GO layers.

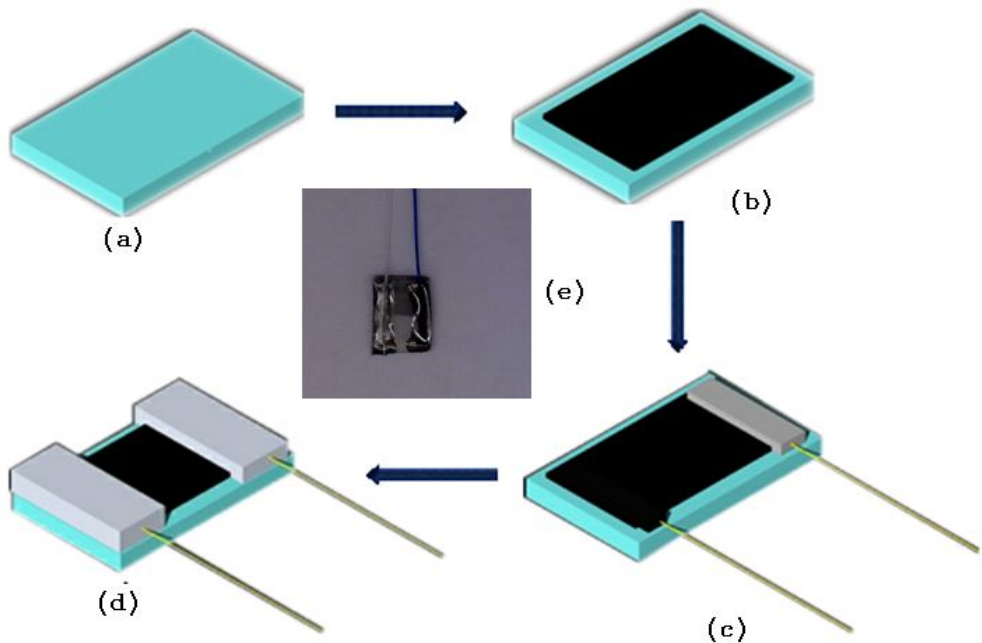

**Fig. 1,** The processes of the fabrication of graphene cell.

The fabrication of the graphene electricity generator was shown in Fig. 1. (a), Glass substrates were cleaned by ultrasonic cleaner in acetone for 10 minutes then in isopropanol for another 10 minutes. Then the substrates were exposed to oxygen plasma cleaner to make the surface hydrophilic; (b), Then the 3% GO solution was spun onto the surface of glass by a spin coater with a speed of 800rpm/min for 40 seconds. Then, the glass was annealed at $250^{o}C$ for 20 mins to reduce the GO film. After the annealing process, the color of the surface of glass changed to gray from light yellow; (c), Silver coated copper wire were adhered to each side of the RGO film by silver paste and graphite paste, respectively; (d), The electrodes were sealed by transparent glue to prevent them from exposing to the solution. One of the samples was shown in Fig. 1 (e).

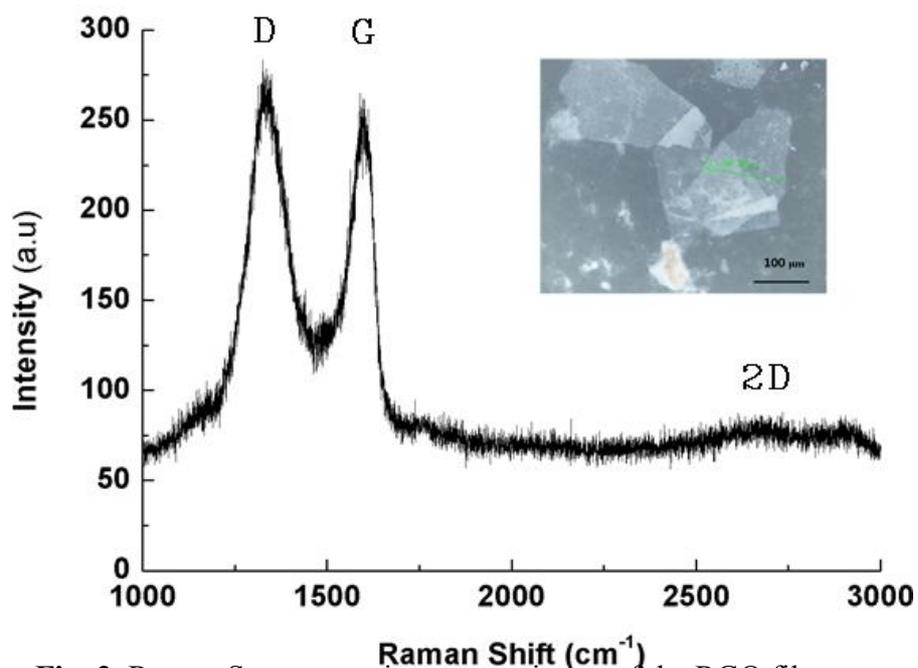

**Fig. 2**, Raman Spectrum microscope picture of the RGO film.

The optical microscope showed that the size of the graphene grain was about 100um (Fig. 2). The Raman Spectrum of the RGO showed that the reducing process was not well performed[5], maybe due to the low temperature and the short time for exposing to the 250°C environment, which limited the performance of the graphene 'battery'.

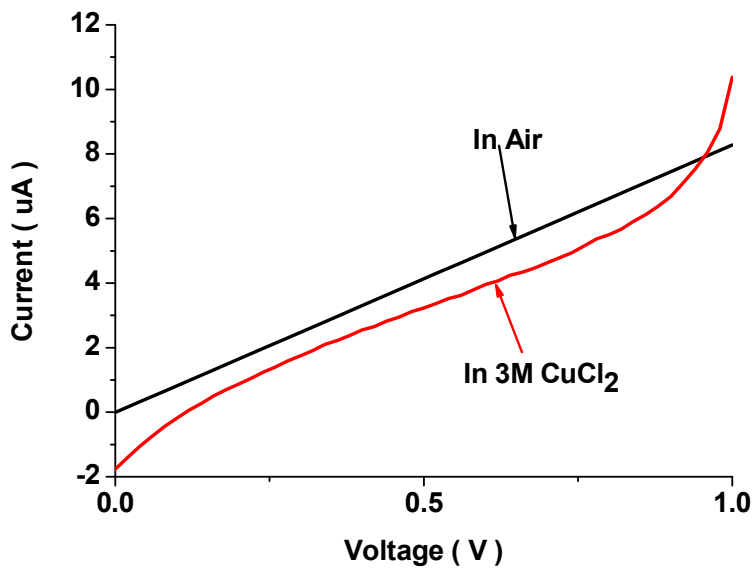

**Fig.3,** I-V curve of the device.

The I-V curve showed that ohmic contact was produced (Fig 3, black). Then the graphene device was packed into a container with 10mL 3M $CuCl_2$ aqueous solution in it. The I-V curve showed that a diode was made when the device was put into the solution (Fig 3, red). A potential difference was made between two electrodes. From the I-V curve, we can find that the open-circuit voltage and the short-circuit current were about 100mv and 1.7uA, respectively.

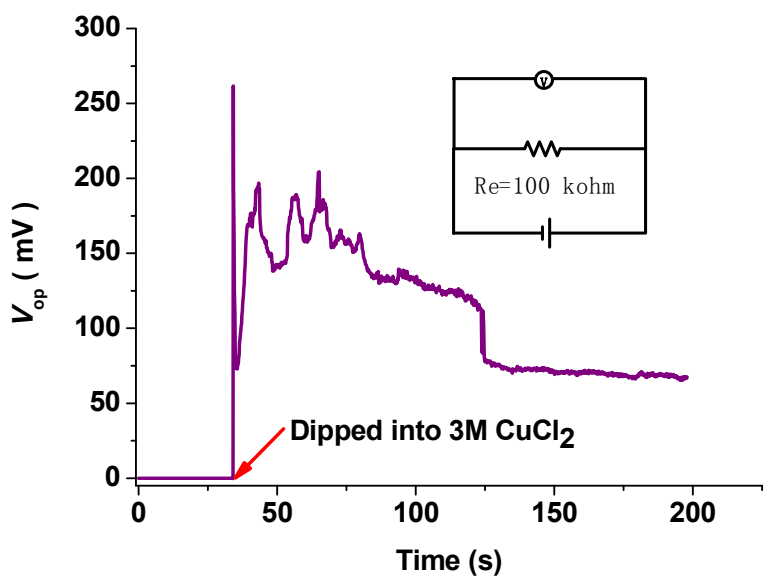

Fig 4, The $V_{op}$-time curve

A 100kohm resistor was loaded to a new device to measure the output of the device (Fig. 4). When there was no solution in the container, the $V_{op}$ was about 100uV, which can be regarded as noise, then 10 mL 3M CuCl2 was dumped into the container, the $V_{op}$ increased to more than 250mV immediately, and drastic fluctuation was observed. After about 2 minutes, the value dropped to about 60mV and became stable at this value. Four other samples were used to repeat this experiment and similar results were got (See support materials).

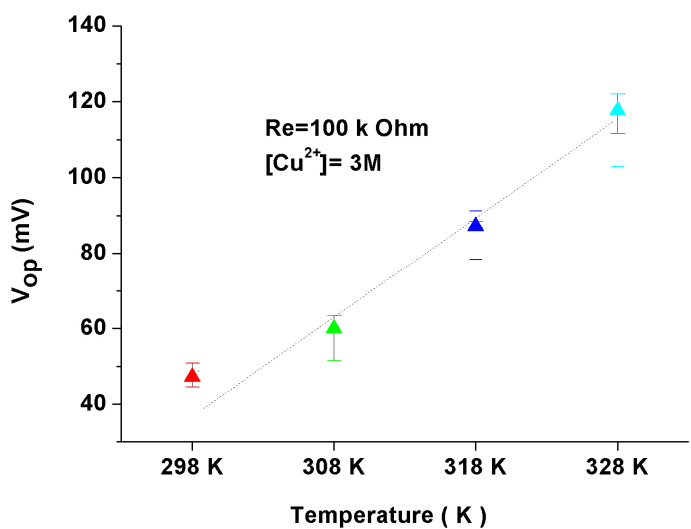

**Fig. 5,** the influence of the temperature on the performance of the device.

In the previous work, we found that the $V_{op}$ showed positive correlation with the temperature [1]. In this experiment, the electricity produced by this system showed positive correlation between $V_{op}$ and temperature, too. The electricity produced by RGO device suffered a bigger fluctuation than the chemical vapor deposition (CVD) graphene device due to the poor quality of the graphene film and the poor contact among the grains of RGO.

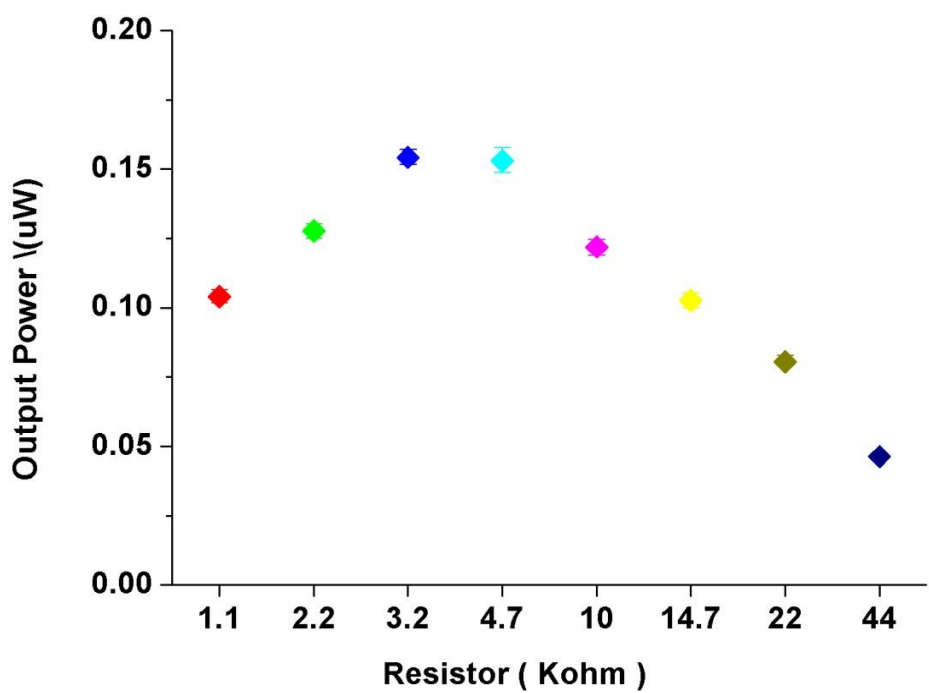

**Fig.6,** The output power of one graphene 'battery'

The output power was tested by loading different resistors to one sample. The results (Fig. 6) showed that the output power reached a peak when a 3.2 K Ohm resistor was loaded to the circuit. The peak power density was about 0.17uW, lower than the value that in the previous work.

In conclusion, this work demonstrated the possibility of producing low cost graphene 'battery' by using low cost reduced graphene oxide, low cost electrodes and substrates. Although the performance of this device was not as good as the device made of CVD graphene used in the previous work, the low cost will make it possible for practical use.And the output power density will be further increased by improving the quality of RGO.

Acknowledge: The author would like to thank Dr. Guoan Tai and Mr. Wenfei Zhang for the discussion of fabrication of graphene oxide, thank Ms. Chang Qin for her help to revise this article, thank Prof. K.H Wong for his supervision during my study in The HKPU.

# Supporting information:

Experimental results with other four samples.

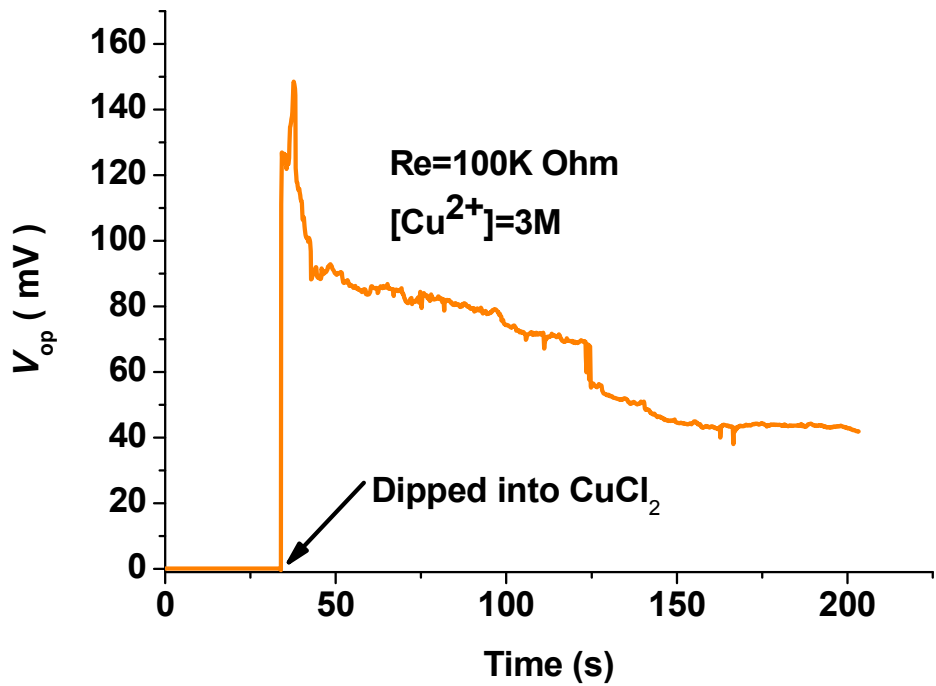

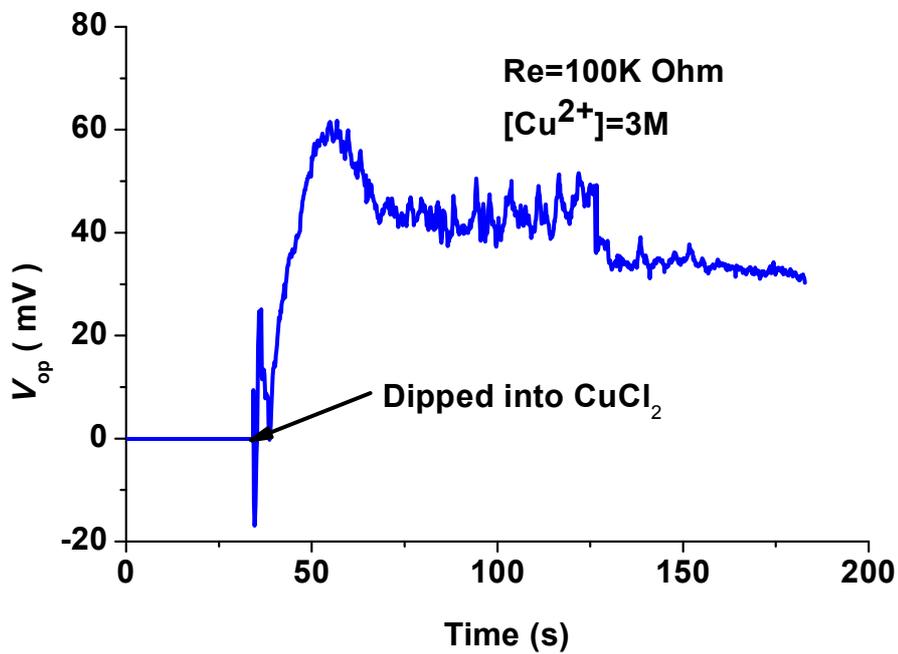

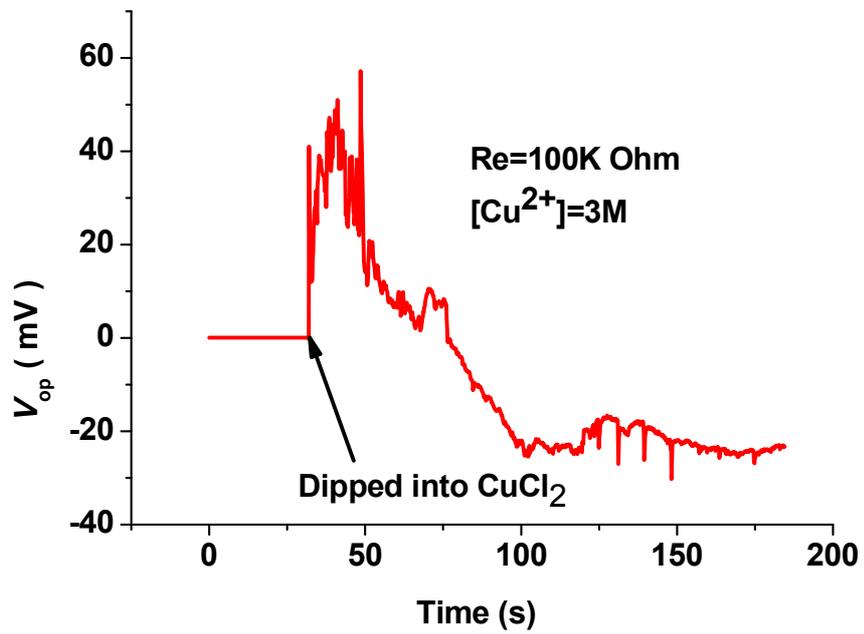

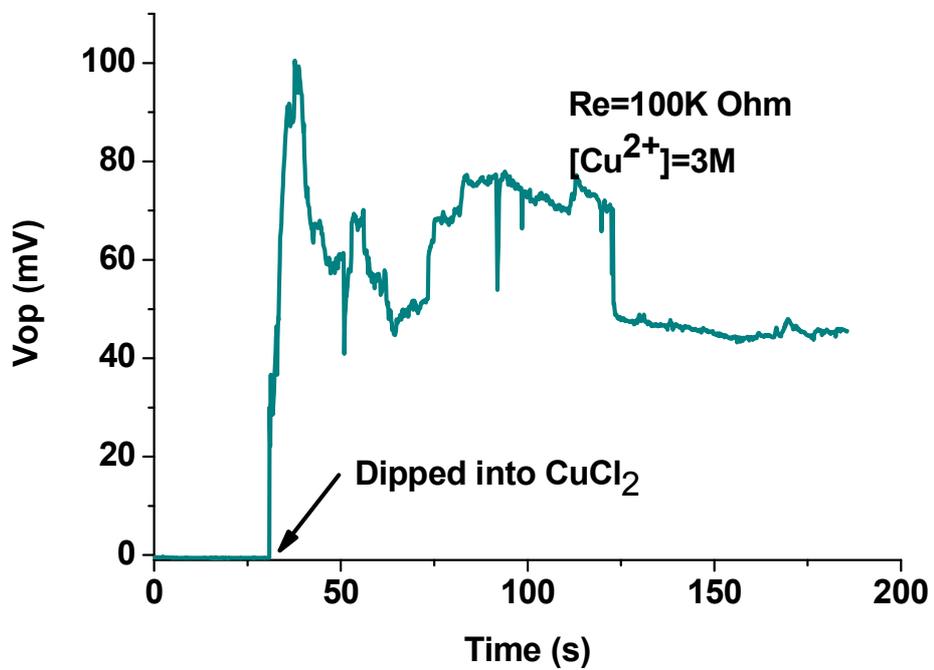